# Examining psychology of science as a potential contributor to science policy

Arash Mousavi [1], Reza Hafezi [2,*], Hasan Ahmadi [3]

[1] Science & Research Policy, National Research for Science Policy (NRISP), Tehran, Iran.

[2] Science & Technology Futures Studies, National Research for Science Policy (NRISP), Tehran, Iran.

[3] Department of Primary Education, Faculty of Psychology and Educational Sciences, Farhangian University, Iran.

* Corresponding author, Address: No. 9, Soheil St., South Shirazi St., Mollasadra Ave., Vanak Sq., Tehran, Iran. Email: hafezi@nrisp.ac.ir

**Abstract**

The psychology of science is the least developed member of the family of science studies. It is growing, however, increasingly into a promising discipline. After a very brief review of this emerging sub-field of psychology, we call for it to be invited into the collection of social sciences that constitute the interdisciplinary field of science policy. Discussing the classic issue of resource allocation, this paper tries to indicate how prolific a new psychological conceptualization of this problem would be. Further, from a psychological perspective, this research will argue in favor of a more realistic conception of science which would be a complement to the existing one in science policy.

## 1. Introduction

The conventional wisdom holds that the science and technology policy is an interdisciplinary enterprise, a crossroad of social sciences. It seems, however, that some disciplines within the family of social sciences have had more opportunities so far for showing their potentials in the field than others. Amongst these lucky disciplines, economics has a noteworthy position. For newcomers to the field (from backgrounds other than economics), there arises almost immediately an impression, amongst the foremost contacts, of this crossroad of ideas: a flavor of 'economics' dominating all around the problems, concerns, methodology, and theoretical contents of the field. This experience turns out to be not so much surprising while considering the fact that a self-organized and rather unbiased sample of scholars working within the discipline contains a majority of near to 60% of researchers with backgrounds in economics (see: (Fagerberg & Verspagen)).

The relative dominance of economics, however, is not bad news for newcomers with other backgrounds. Instead, it shows a vast expanse of virgin potentialities for other disciplines to play their own game within the field. The recent history, indeed, has recorded good examples of the revelation of such potentialities. Appreciating the prolific role that Keith Pavitt played in the field, Daniele Archibugi reminds us that:

> "Some of the fundamental contributions to our understanding of innovation have come from scholars who, like Pavitt, have no formal training in economics. Economics needed to import fresh blood from other disciplines such as engineering, management sciences, natural sciences, history and philosophy of science and knowledge to understand the determinants and impact of technological change" (Archibugi, 2001).

By importing this fresh blood from other disciplines, the traditional issues in science and technology policy can be reformulated in new ways. This can help in finding novel solutions for them. New perspectives can also create their own set of issues and problems. In addition, there is always the possibility of introducing new conceptual and theoretical frameworks applying insights from different disciplines into the field.

For centuries, formal rules were identified based on available information, while in the information era, scientists should cope with information overload. In the past, the focal question was "how to collect data" and now the main challenge has been changed to "how to manage/ analyze data". Then psychology of science attempts to uncover how scientists' minds are adapting to a new world with too much information (Webster, 2012).

Before calling it "psychology of science", and recognizing itself as an independent scientific branch, other researchers showed their interests in decoding intelligence, cognition, and personality as the main targets of the psychology of science. For example, Eiduson and Beckman (Ediuson & Beckman, 1973) analyzed scientists compared to other people in terms of personality, demographics, and biological traits. However, findings of such studies were not reliable and existence of obvious differences are hardly accepted.

Archibugi's list of examples includes most of the well-known contributing disciplines (Archibugi, 2001). It lacks though some of the less-familiar enormously potential fields. In what follows, we will examine one of these disciplines, a newly emerging sub-field of psychology, the psychology of science, to see how fruitful would be to invite it into the family of social sciences which all contribute to the interdisciplinary field of science policy. We exclude technology policy and issues relating to scientists in the industry because psychological research on these topics is still scarce.

After a brief introduction to the psychology of science, we will try to bring into view samples of traditional science policy problems by looking at them from a psychological perspective. Applying some psychological techniques, our challenge will be to find out whether these problems can be addressed in more productive ways. Last but not least, we will seek to propose some unprecedented issues that can be raised from this new psychological point of view and may help to a more comprehensive science policy.

## 2. The Psychology of Science: a Nascent Discipline

Within the family of science studies (meta-sciences), psychology of science is the youngest member. Far behind such fully established disciplines as philosophy, sociology, and history of science, the psychology of science is still in its earlier stages of development. It does not enjoy yet Joseph Matarazzo's criteria for the establishment of a new field (i.e., its own association or society, journal, postdoctoral training programs, etc.) ([Stone, Weiss, Matarazzo, Miller, & Rodin, 1987](#)). Amongst the reasons for this comparative delay, two may be more important. First, the long-lasting image of scientists as some species of comprehensive super-humans with no passions or emotions meddling into their timeless process of self-sufficient rationality did not leave many open doors for psychologists to consider themselves as relevant components in the studies on science. Even these allegedly pure activities of science (discovery, theory evaluation, and so on) have been viewed until relatively recently as exclusive subjects of normative investigations of 'logicians' rather than descriptive issues to be studied by psychologists (see ([Thagard, 1993](#))). Second, even now much of the psychology of science is dormant, latent, and implicit. Indeed, there are more than a few talented minds in psychology who are conducting research on scientific thought, behavior, interest, talent, and creativity. They just do not identify themselves with or are not aware of the term 'psychology of science' ([G. J. Feist, 2006b](#)).

For about 30 years, since the end of the 1930s when the psychology of science was born officially, it has experienced a period of silence ([Yanhui & Jianshan, 2019](#)). It has become popular in the 1980s represented by ([Brannigan & Wanner, 1983](#); [Campbell, 1982](#); [Grover, 1981](#); [Simonton, 1988](#); [Tweney, Doherty, & Mynatt, 1981](#)). Psychology of Science Conference was held in 1986 at Memphis State University where research groups were formed and the basis of the psychology of science was discussed. After 2000, "the psychology of science and the origins of the scientific mind" by Feist ([G. J. Feist, 2008](#)), and "psychology of science: Implicit and explicit processes" edited by Proctor and Capaldi ([Proctor & Capaldi, 2012](#)) concepts and directions of this new member of philosophy of science were cleared.

Among the very rare reviews of the discipline, is the work done by Gregory Feist and Michael Gorman ([G. J. Feist & Gorman, 1998](#)) (revised and extended in ([G. J. Feist, 2008](#))) in which they propose an integrative and organizing model for the psychology of science. This model (see figure 1) summarizes the main factors that lie at the foundation of scientific interest, talent, and achievement. The circles in the model indicate the five major domains in which the current literature is developing: biological, developmental, cognitive, personality, and social psychology of science. The size of this essay does not permit us to provide a detailed survey of each of these sub-disciplines. For each of them, therefore, it may suffice to identify the most important issues and problems and some samples of its findings. This section draws mainly upon Gregory Feist's works ([G. J. Feist, 2006a](#), [2006b](#), [2006c](#), [2008](#); [G. J. Feist & Gorman, 1998](#))

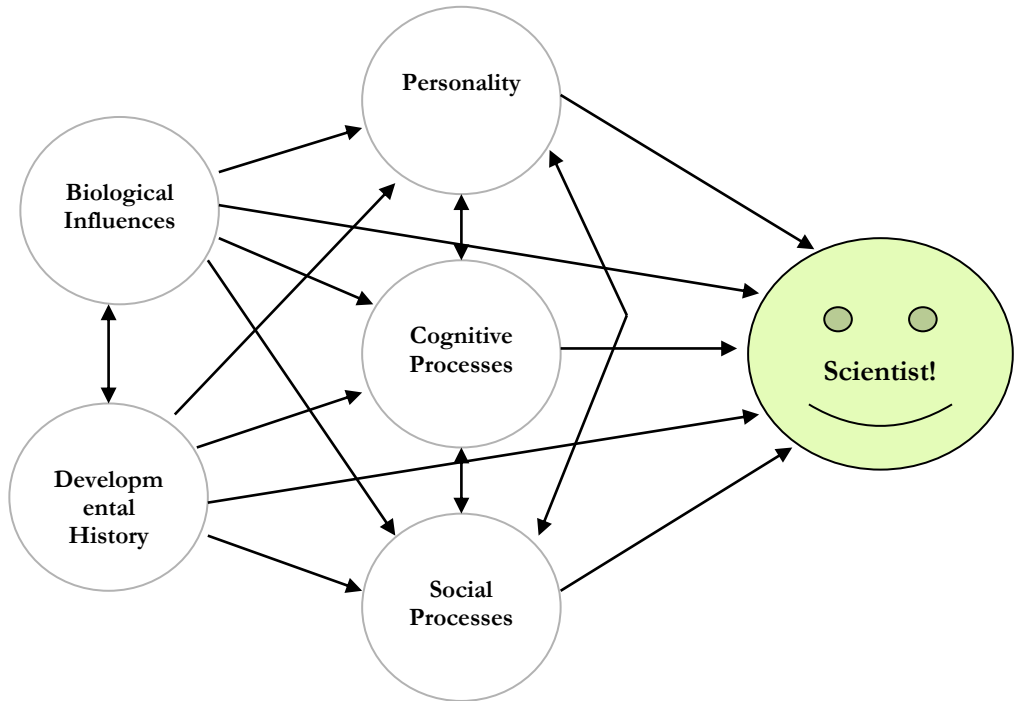

**Figure1. The Structure of Contemporary Psychology of Science** (G. J. Feist & Gorman, 1998)

● *Biological psychology of science*   One of the most interesting issues in psychology of science concerns the biological and genetic roots of scientific talent. Lots of efforts have been devoted to this area to find out whether some unique configurations of genetic factors can explain mathematical genius and creativity. There is also a lively discourse on the role gender plays in science: Are there differences between males and females in mathematical ability? Do men and women produce scientific works at different rates? Is there a gender difference in the quality of these works?

● *Developmental psychology of science*   Psychologists of science has been curious for a long time about the shape of the diagram which indicates the relationship between scientific productivity and age. Feist and Gorman's conclusion is that a consensus exists over this issue: "there is a curvilinear relationship between age and productivity, with the peak generally occurring in one's late 30's or early 40's" (G. J. Feist & Gorman, 1998). Some other problems in this domain include: Does producing works early to predict later levels of productivity? Are older scientists more resistant to scientific revolutions than younger ones? What role do family members or teachers play in promoting scientific interest? Does being trained by an eminent scientist predict obtained eminence? What role does birth order or religious background play in scientific success or interest?

● *Cognitive psychology of science*   Most of the research done in biological and developmental psychology of science have been essential of statistical nature. To become a well-developed and paradigmatic discipline, however, psychology of science needs to foster conceptual frameworks and theories of its own. For this purpose, the cognitive psychology of science has already revealed itself as the most promising area within the discipline. Relying on epistemological insights provided by philosophers of science and also taking advantage of computational terminology and techniques supplied by the researchers working in the field of artificial intelligence, cognitive psychologists of science have begun creating testable models which all try to simulate the basic scientific tasks. These tasks, as Brewer and Mishra suggest, are generally of three kinds: (a) Understanding and evaluating scientific information; (b) Generating new scientific knowledge, and; (c) Disseminating scientific knowledge (Bechtel, Graham, & Balota, 1998).

● *Personality psychology of science*   Four fundamental topics in this area of research are (a) Consistent personality differences between scientists and non-scientists; (b) Consistent personality differences between eminent and less eminent scientists; (c) Consistent personality differences among scientists of different theoretical persuasions; and finally (d) The directional influence of personality on scientific behavior. Empirical literature over the last 60 years, for example, have converged on a description of scientists as "more conscientious, driven, introverted, emotionally stable, and controlled compared with non-scientists" (G. J. Feist & Gorman, 1998).

● *Social psychology of science*   Gordon Allport, the renowned American psychologist, defines social psychology as "an attempt to understand and explain how the thought, feeling, and behavior of individuals are influenced by the actual, imagined or implied presence of others" (Allport, 1985). As we are increasingly getting aware of non-cognitive and highly social aspects of scientific practice, the social psychology of science is becoming more relevant as a useful approach to the study of scientists. This sub-discipline has so far provided us with insightful explanations for how new ideas develop, are communicated, are evaluated, and become pervasive within a group of scientists. Applying the well-established theories of social psychology, researchers have been led to deeper understandings concerning such issues as the tensions between orthodoxy and heterodoxy, the issue of peer review and quality monitoring in science, citation patterns, and scientific teamwork. These are, however, only a small proportion of potentialities inherent in the social psychology of science as well as the psychology of science as a whole.

● *Scientific reasoning*   Still the concept of "scientific reasoning" is an open question among researchers. Not that there is no consensus on it, but since scientific reasoning has been studied from different perspectives by various researchers from different scientific fields. The concept of scientific reasoning can be studied from the psychological and sociological viewpoints, to investigate how it defines through the lens of psychology of science.
In simple words, the reasoning is defined as an explanation of a phenomenon that is focal and desired goal of scientific inquiry. From the psychological viewpoint, as Zimmerman noted, causal reasoning and explanation reflect this conceptualization (Koslowski, 2013; Zimmerman, 2000, 2007). For example, the explanation is reflected in the study of an event to identify it is the cause.

As this paper aims to review and summarize existing literature, we distinguish some main views of scientific reasoning and explanations from the psychology of science perspective. These include the following:

1. To detect causal mechanisms: these types of studies focused on a research question that examines the causal event in a specified situation. For example, in such studies, researchers attempt to identify the causal relationship among event X and event Y based on Humean indices such as priority, contiguity, and covariation (Macnabb, 2019). Some of such researches are investigated by (Gopnik, Schulz, & Schulz, 2007; Haith & Benson, 1998; Koslowski & Masnick, 2002; Koslowski & Thompson, 2002; D. Kuhn et al., 1988; Schulz & Gopnik, 2004).

2. The importance of background information: as Koslowski (Koslowski, 2013) noted, applying identified formal rules will fail unless background information about the explanation and causal mechanism is taken into account. This issue points out that scientists formed explanations based on relevant alternatives, rather than all possible alternatives (Boyd, 1989; Darden, 2006; Fine, 1984; Koslowski, 2013; T. S. Kuhn, 2012; Lipton). Note that, specifically, formed rules from a psychological perspective are characterized differently compared to what is known as the mean indices (for more information about psychological formal rules see: (Bonawitz & Lombrozo, 2012; Lombrozo, 2007; Samarapungavan, 1992).

3. The experiment of science differs in theory and practice: formal rules cannot support perfect science, so the philosophy of science proposed descriptions of scientific practice-based background information. Inferencing to be the best explanation is an accepted criterion to describe a perfect scientific practice which argues that the "explanation" will be agreed upon if it overcame competitive alternative explanations since it provides more logical and detailed causal relations among others (Harman, 1965; Lipton, 2003; Magnani, 2011; Proctor & Capaldi, 2006). In this manner, plausibility is a key factor to accept the best explanation.

4. In practice, science tends to make good outcomes rather than to guarantee them: in many scientific fields, actual relevant background information does not exist to form an alternative hypothesis. In other words, they might not have been discovered/ proved/ obtained. For example, physicists have not proposed a comprehensive model to explain quantum mechanics perfectly yet, or currently, definitive treatment for the COVID-19 pandemic is not proposed. In such circumstances, explanations are accepted since they make good outcomes, rather than to guarantee them. Consistency is a crucial key index to accept or reject a formal rule.

*scientific personality*   One of the many building blocks of scientific thought and behavior is personality. Career interests in general and scientific career interest and talent, in particular, stem from personality and individual differences in thought and behavior (Feist,2013).

What does *personality* mean? Individual differences exist in the way people think about past changes to their personality traits (Cochran & Haas, 2020). Personality is that pattern of characteristic thoughts, feelings, and behaviors that distinguishes one person from another and that persists over time and situation (Heinström, 2013). Personality is a ―pattern of relatively permanent traits and unique characteristics that give both consistency and individuality to a person's behavior (J. Feist & Feist, 2009). Personality influences how people interact with their environment and interpret the particular meaning of the situations created by the environment (John, Naumann, & Soto, 2008).

Personality traits exist as a multileveled hierarchical structure and are relatively stable over the course of life (Reason, 1994). After 50 years of personality research, there is a common agreement in the field that there are five basic dimensions that can be used to describe differences in cognitive, affective, and social behavior. This is the base for the five-factor model of personality (Heinström, 2013). The five dimensions are depicted in Table 1.

Table 1. Personality dimensions and the poles of traits they form (Heinström, 2013)

| Personality dimension | High level | Low level |
|---|---|---|
| Neuroticism | sensitive, nervous | secure, confident |
| Extraversion | outgoing, energetic | shy, withdrawn |
| Openness to experience | inventive, curious | cautious, conservative |
| Agreeableness | friendly, compassionate | competitive, outspoken |
| Conscientiousness | efficient, organized | easy-going, careless |

The Five-Factor Model (FFM) comprises five bipolar factors: openness (imaginative – down-to-earth, (conscientiousness (well organized – disorganized), extraversion (outgoing – reserved ,( agreeableness (trusting – suspicious), and neuroticism (anxious – calm), which are placed on the continuum. All five factors are distributed normally across the population (G. J. Feist, 2008).

*Co-authorship*   Science and technology policy academics and evaluators use co-authorship as a proxy for research collaboration despite knowing better. We know better because anecdotally we understand that an individual might be listed as an author on a particular publication for numerous reasons other than research collaboration. Yet because of the accessibility and other advantages of bibliometric data, co-authorship is continuously used as a proxy for research collaboration (Ponomariov & Boardman, 2016).
 In recent decades there has been growing interest in the nature and scale of scientific collaboration. Studies into co-authorship have taken two different approaches. The first one attempts to analyze the reasons why authors collaborate and the consequences of such a decision.

The second approach is based on the idea that co-authorship creates a social network of researchers. In recent years, the collaboration between scientists increased. This collaboration can be formal (joint papers, the guidance of doctoral dissertations, and participation in research groups) or informal (arising from the comments of colleagues, reviewers, and editors) (Acedo, Barroso, Casanueva, & Galán, 2006).

*Human research participation* Research refers to a class of scientific activities designed to develop or contribute to generalizable knowledge. The term "human research", then, refers to research that involves human subjects. Research is a public trust that must be ethically conducted, trustworthy, and socially responsible if the results are to be valuable (Fakruddin et al., 2013).
Over the past few decades, academic discussions in the broad contexts of public engagement in science policy, discourse ,and research have taken a "participatory turn" (Jasanoff, 2005). `For the last 15 years, Peter Reason has been developing a democratic model of research where researchers organize a group of people to study a phenomenon of concern to them, regardless of their educational background. Labeled `human inquiry′, we have always had a more theoretical justification of the process than actual description (Reason, 1994).

## 3. A Psychological Perspective on Science Policy
Browsing the existing literature on science policy, one cannot resist Lundvall and Borras's conclusion that "the major issues in science policy are about allocating sufficient resources to science, to distribute them wisely between activities, to make sure that resources are used efficiently and contribute to social welfare" (Lundvall & Borrás, 2005). These issues are all of the economic characters. In fact, these issues are the main focus of a relatively new branch of economics, namely the 'economics of science.' Introducing this sub-field of economics, it is interesting that, Ed. Steinmueller inscribes almost the same issues that Lundvall and Borras ((Lundvall & Borrás, 2005)) ascribe to science policy. He writes: "Determining the principles governing the allocation of resources to science as well as the management and consequences of the use of these resources are the central issues of the economics of science" (Pavitt, Steinmueller, Calvert, & Martin, 2000). They correspond with two simple but fundamental questions:
    (1) The question of *why*, i.e., why do we fund science?
    (2) The question of *how*, i.e., how do (or should) we fund science?[1]
The most noteworthy answer to the first question has been provided by the 'simple economics' of Richard Nelson (Nelson, 1959) and Kenneth Arrow (Arrow, 1972). Science, they argue, possesses a set of properties which is adequate for considering it to be an instance of the 'public goods.' Like any other public commodity, it faces, therefore, the challenge of 'appropriability' and needs governmental support.

---

[1] A third question may be put forward as *how much* should we finance the science as compared with other public sectors? The question has received a good deal of economists' attentions. We do not have enough space to discuss it here though.

The second question though has not been as straightforward to answer as the first one. Nor it has been possible to find a purely economic answer for it. Economists, in this case, owe a substantial debt to a sociologist, Robert Merton, for providing them with a good starting point. Science, as Merton demonstrates, is a highly competitive contest over the goal of *priority of discovery*. One who scores in this game is rewarded in varied forms by the scientific community (Merton in a series of articles; cited in (Stephan, 1996)) and by the broader society as well. Dasgupta and David were the first to realize the economic importance of this reward system (Dasgupta & David, 1987). While the scientists are driven and occupied by this non-market-based race over incentives, they are serving the economic requirement to the public disclosure of knowledge (Pavitt et al., 2000; Stephan, 1996). Priority, therefore, is (and ought to be) a good criterion for the allocation of resources to science.

These are not the whole work done in response to the main challenges of science policy. Nonetheless, they may represent the sort of rationality that is dominant in the field and pave the way for us to return to our primary question in this essay, i.e., how can a psychological perspective contribute to reformulating or even finding novel solutions for the main problems in science policy?

The mainstream economic view of science is simple. In this view, the market is the general paradigm for all modern social organizations and science is just one special case of this generic structure, a marketplace of ideas (see: (Mirowski & Sent, 2002)). The internal logic of the production of scientific knowledge, as we saw above, may be different from other market-based institutions, but the result from an economic point of view on resource allocation is the same: those who produce more receive more. Scientific 'productivity' is the most important variable in this view and its measurement is critical for policy decisions over resource allocation.

From a psychological point of view, scientific 'productivity' is not a 'given' variable which should exclusively determine policy decisions. It is, as we saw throughout section 2 above, a function of at least five categories of psychological variables. A scientist's level of productivity is determined by her biological and developmental history, by her cognitive capacities, by her personality, and by the people who are present around her. Even if we are restricted within the single goal of increasing scientific productivity and even if we have in hand the single tool of fund allocation, there are many more opportunities for us to expend our money than simply enriching the strongest links.

Biological psychologists of science are able, for example, to show how effective would be for the mathematical productivity of the next female generation investing in a campaign against the established stereotype of 'mathematics as a masculine enterprise' (see (Benbow, 1988)). Developmental psychologists of science can help policymakers to realize the immense importance of subsidizing the popularization of characters like John Nash in movies like 'The Beautiful Mind' ((Nasar, 2011)) for the process of example-making of our teenagers and, therefore, for the aggregate scientific productivity of the next generation. Cognitive psychologists have considerable capability to provide science policy with programs that would enhance scientists' self-consciousness while performing their cognitive tasks. Personality psychologists can perform a complementing role in employment filter mechanisms in science which has traditionally made schooling success the only indicator of the potential for scientific contribution (see (Pavitt et al., 2000)). Finally, social psychologists of science can adjust policy maker's understanding of the very measurements of productivity by providing complicated and systematic analyses of publishing and citation patterns.

Our list of examples can go on to include all psychological determinants (positive and negative) of scientific productivity. The underlying logic though is quite simple. A funding regime based on merely 'economic' considerations is a 'posterior' regime. It takes the *status quo* of scientific production as given and tries to retain and improve it. This system of science governance has lots of shortcomings. It bears, for example, a sort of Matthew effect which diminishes diversity in science and leaves few resources for important scientific activities other than 'production' such as disseminating scientific results (Dasgupta & David, 1987). A psychological approach can make this regime more complete by adding to it an 'anterior' perspective which creates the possibility of enhancing productivity by manipulating its determinants.

**4. Beyond the Economic Paradigm**

There is more to psychology of science than focusing only on scientific productivity. Scientific production is important but just one aspect of scientific life. There is more also for science policy to be concerned about besides merely economic issues like resource allocation. Economics, by its nature, makes a thing-like picture of science, a commodity, which in turn results from a production process conducted by some semi-robotic species called a scientist. This is useful but not the whole picture. Mirowski and Sent are right when they write:

> "It is a commonplace observation that economics love the individual; it is just real people that they cannot be bothered about. A wag once added that economists also profess to love Science; it is just *real scientists* that make them nervous" (Mirowski & Sent, 2002)

It is precisely the life-world, *Lebenswelt*, of these *real scientists* that need to be studied from within. Science, as a unique sort of social life with its own set of shared practices, beliefs, values, institutions, and structures of interaction, requires researchers who are studying it to do more than reporting raw data. To continue to live healthy, this complicated world calls for policymakers to enhance their understanding of the very notion of 'the health of science.' The psychology of science along with her methodological complexity (including both objective and hermeneutic methods) and also novel and promising theories[1] can play a great part in this process.

---

[1] I have in my mind, particularly, the 'Role Theory' within the subfield of social psychology. It takes another essay though to develop this idea.

## 5. Directions of the psychology of science:

As noted by Gregory D. Webster, in the first decade of the 21st-century psychology of science was targeted by researchers in ascending order. Figure 2 schematically presents results for changes in google scholar hit counts from 2000 to 2009 (Webster, 2012).

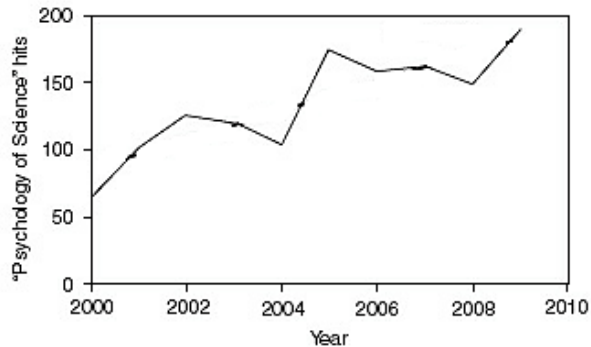

**Figure2. Results for google scholar hits for "psychology of science" from 2000 to 2009 (Webster, 2012)**

As presented by figure 2, publications and citations related to "psychology of science" rose from 66 in 2000 to 189 in 2009. Moreover, linear regression analysis showed that the rising trend was largely linear (Webster, 2012).

To dedicate an updated and comprehensive outlook we have investigated a decade from 2010 to 2019 based on the google scholar database. The search strategy was to recall publications which exactly pointed to "psychology of science" in their titles. After a refining process to eliminate outliers, 32 items were detected. Figure 3 represents publication frequency over time and by type from 2010 to 2019.

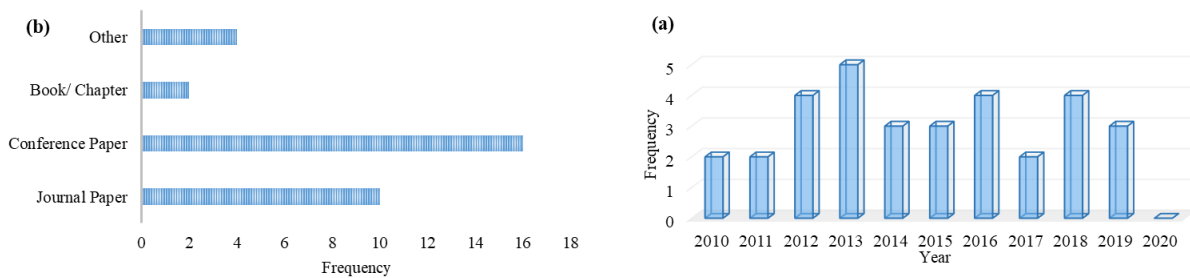

**Figure3. Frequency of published documents (a) by time (2010-2019); (b) by types**

The review revealed that still psychology of science needs more attention and it is growing slowly. Moreover, a significant portion of published works (in the last decade) were books (10 out of 32) which show that there are still debates about focal concepts and theoretical foundations. In addition, a new research line was observed which is called "social psychology of science". The social psychology of science is aimed at investigating social aspects of scientific reasoning, personality, etc., rather to study scientists as individuals. The core idea is that social factors influence researchers dramatically. Some examples are: ([Holtz, 2016](#); [Johnson, 2018](#); [Paletz](#); [Purkhardt, 2015](#)).

**7. Conclusion**
As a subfield of psychology, the psychology of science is still in its infancy. It is growing rapidly though and this rapid growth, the essay suggests, is providing a unique opportunity for science policy-makers and analysts to expand their understanding of science toward new horizons. The management of aggregate scientific behavior of a nation, including quantity and quality of scientific production, can be improved by adding to the knowledge base of managers a systematic comprehension of biological, developmental, cognitive, personality, and social variables that influence scientific behavior. The introspective and participation methods in psychology can add to the statistical and objective investigations of science the possibility of understanding and interpreting the internal environment of science as it is. This equips policymakers with some sort of conservatism which may be useful against periodical waves of radical change.

From the meta-science viewpoint, psychology of science is totally different from the classic psychology of science. Psychology of science as an emerging branch of meta-science aimed to use psychological research to uncover how scientists create explanations and to study their characteristics in terms of cognition and personality especially compared to non-scientists members of the society. In the 21$^{st}$-century, psychology of science distinguished itself from the classic psychology of science and developed emerging branches such as "social psychology of science".

The social psychology of science challenged standard approaches as they attempt to analyze scientists in terms of cognition, personality, intelligence, etc. individually, while social interactions, roles, and cultural factors are ignored.

Psychology of science has not been yet recognized as a formal discipline in developing societies. Although scholars from philosophy of science, psychology, sociology, etc. majors conducted some attempts, however, they are far from idealized psychology of science research. Still there are limitations to develop psychology of science in developing societies. First, and almost the most important one, the discourse of psychology of science has not been formed and related concepts are not cleared. Second, as a developing field, it needs particular research methods, at least to some extent. Third, the scientific community of psychology of science consists of a multi-disciplinary approach that includes disciplines such as philosophy of science, psychology, sociology, and science policymaking to form an emerging inter-disciplinary research field. There are really few scholars who have sufficient knowledge about all of these majors. Our initial recommendation is to establish an independent scientific journal to create forums for exchange of views and initialize psychology of science scientific community.

**Ethical statement:**

This article does not contain any studies with human participants (i.e. questionaries, survays, interviews etc.) or animals performed by any of the authors.